\documentclass[11pt]{article}
\usepackage{moriond,epsfig}

\def\be{\begin{equation}}
\def\ee{\end{equation}}
\def\bea{\begin{eqnarray}}
\def\eea{\end{eqnarray}}

\begin{document}
\vspace*{4cm}
\title{MEASUREMENTS OF $|V_{ub}|$ WITH SIMULTANEOUS REQUIREMENTS ON $M_X$ AND $q^2$ BY BELLE}
\author{ T. NOZAKI}
\address{High Energy Accelerator Research Organization (KEK), Oho 1-1, Tsukuba, Ibaraki, Japan
}
\maketitle
\begin{figure}[h]
\begin{center}
\includegraphics[height=3.5cm]{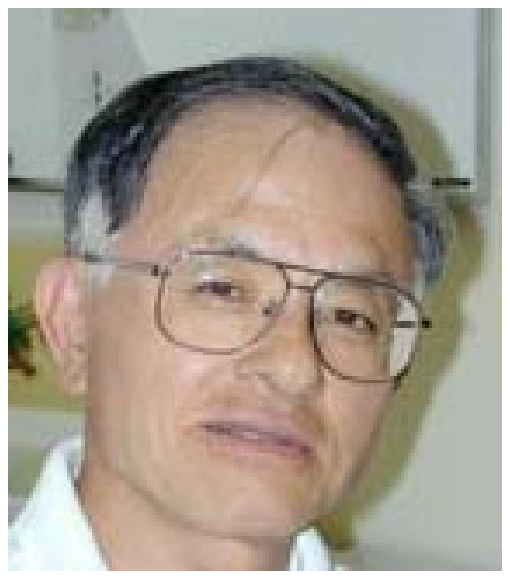}
\end{center}
\end{figure}
\abstracts{
We report the measurement of an inclusive partial branching fraction for charmless semileptonic $B$ decay
and the extraction of $|V_{ub}|$.
Candidates for $B\rightarrow X_u \ell \nu$ are identified using a novel $X_u$-reconstruction method with simultaneous requirements on the hadronic recoil mass ($M_X$) and the invariant mass squared of lepton-neutrino system ($q^2$).
Based on 86.9 ${\rm fb}^{-1}$ of data taken with the Belle detector, we obtain
${\Delta \cal B}(B \rightarrow X_u \ell \nu ; M_X<1.7\ \mathrm{GeV}/c^2, q^2>8.0\ \mathrm{GeV}^2/c^2)  = (7.37 \pm 0.89(\mathrm{stat.}) \pm 1.12(\mathrm{syst.}) \pm 0.55(b \rightarrow c) \pm 0.24(b\rightarrow u) ) \times 10^{-4}$
and determine 
$|V_{ub}| = (4.66 \pm 0.28(\mathrm{stat.}) \pm 0.35(\mathrm{syst.}) 
\pm 0.17(b\rightarrow c)\pm 0.08(b\rightarrow u)
 \pm 0.58(\mathrm{theory}))\times 10^{-3}$. }

\section{Introduction}
%\subsection{Producing the Hard Copy}\label{subsec:prod}

\begin{figure}[htb]
\includegraphics[width=8.0cm,height=5.5cm]{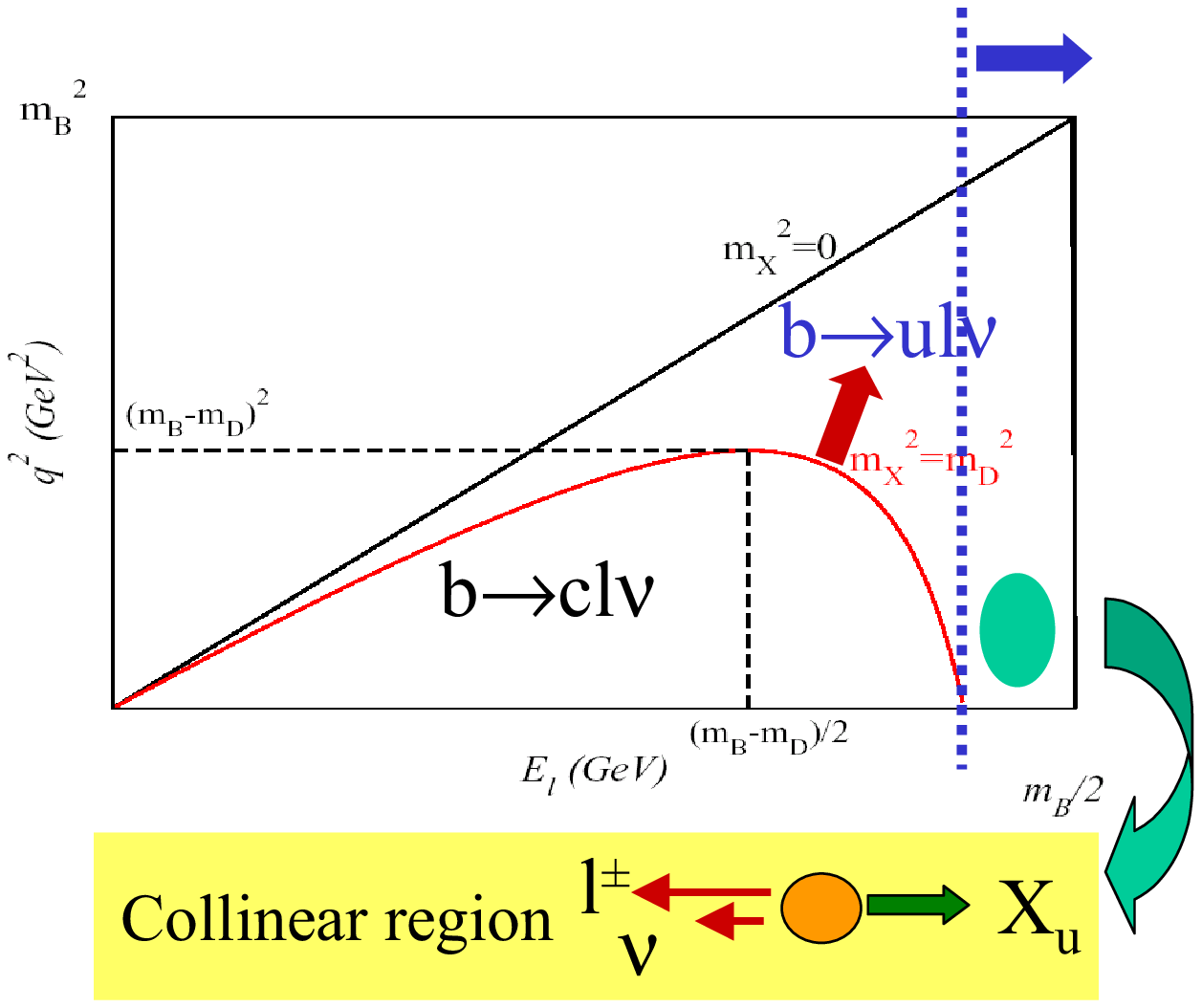}
\includegraphics[width=8.0cm,height=5.5cm]{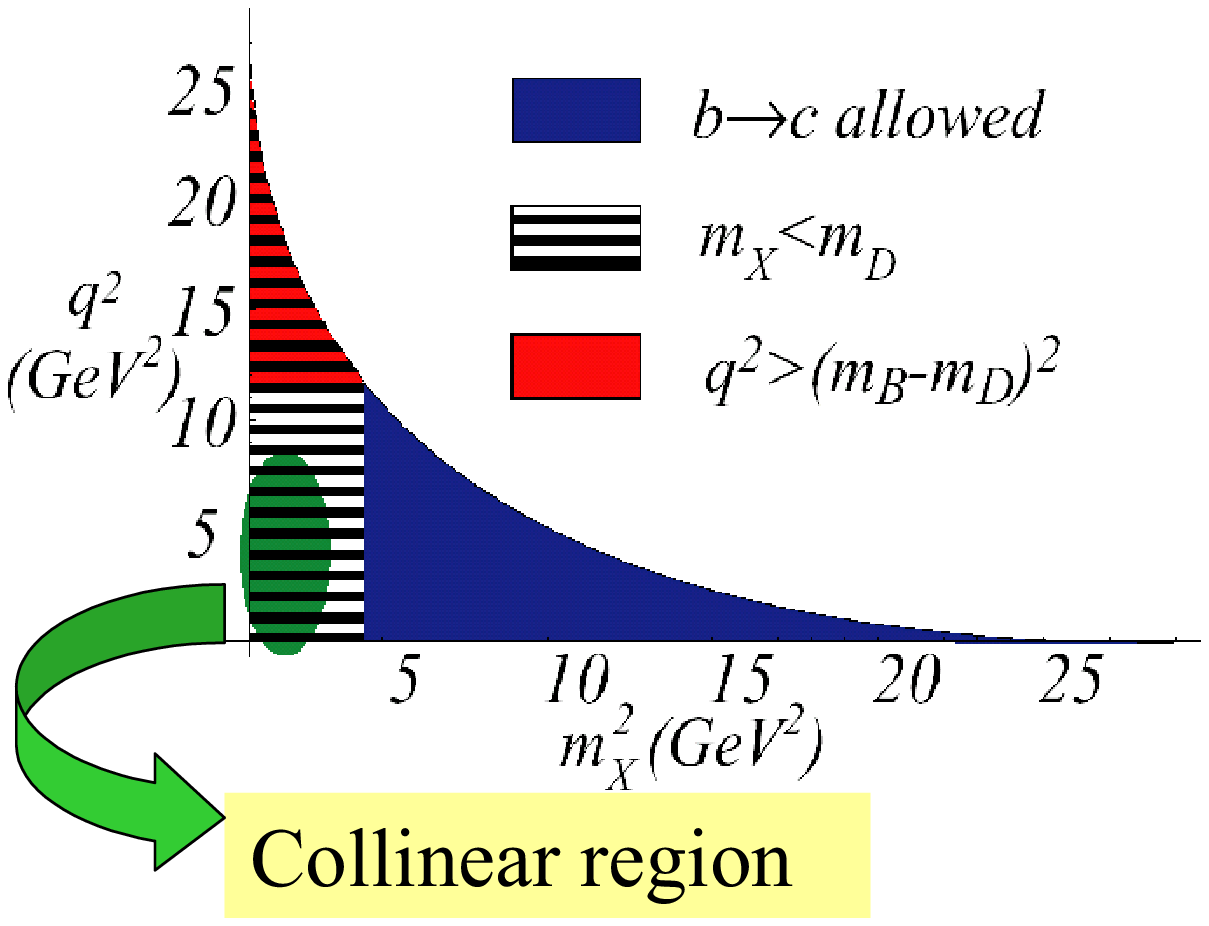}
\caption{Reduction of charmed events by $E_l$ or $M_X$ cuts and definition of the collinear region.}
\label{elmx}
\end{figure}

The off-diagonal element $V_{ub}$ in the CKM matrix is an important ingredient in overconstraining the unitarity triangle by measuring its sides and angles.
The inclusive charmless semileptonic decay $B\rightarrow X_u \ell \nu$ is the most promising process to determine $|V_{ub}|$ since $|V_{ub}|$ is related to the total branching fraction of  $B\rightarrow X_u \ell \nu$  model independently by OPE and HQET with a precision of about 7$\%$\cite{PDG2002} as follows.
\begin{equation}  
|V_{ub}| = const \times \sqrt {\mathcal{B}(B \rightarrow X_u \ell \nu )}
                       (1\pm0.07(theory))
\label{eq:pdgvub}
\end{equation}
Experimentally we have to apply a cut on the phase space  to suppress
the hundred times larger contribution of background from $B\rightarrow X_c \ell \nu$.
The  experiments on the $\Upsilon(4S)$ resonance have used  the limited  region of lepton momentum $E_l$~\cite{CLEOendpt}  or the hadronic recoil mass $M_X$~\cite{BaBar} as seen in Figure~\ref{elmx}.  The phase space of these experiments contains the so-called collinear region shown in Figure~\ref{elmx}. In this region OPE fails and  a shape function of the B meson is required.
Therefore the need to extrapolate measured rates from such limited regions results in large theoretical uncertainties on $|V_{ub}|$. 
A recent theoretical development suggests that one can significantly reduce the theoretical uncertainty on the extrapolation by applying simultaneous cuts
on  $M_X$ and the invariant mass squared of the lepton-neutrino system ($q^2$)~\cite{bau}. 
We compare the extrapolation errors for three cases assuming  
$m_b^{\rm{1S}}=4.70\pm0.12\,{\rm{GeV}}/c^2$ where $m_b^{\rm{1S}}$ is one-half of the perturbative contribution to the mass of the $\Upsilon(1S)$.
Case 1 ($E_l > 2.2\ \rm{GeV}$):20$\%$; case 2 ($M_X <1.55 {\rm{GeV}}/c^2$):18$\%$; case 3 ($M_X <1.7\ {\rm{GeV}}/c^2$ and $q^2 > 8.0\  {\rm{GeV^2}}/c^2$):10$\%$.
There are three methods to reconstruct $M_X$ and $q^2$. The first one is a traditional neutrino reconstruction method used by CLEO which reconstructs the neutrino 4-momentum from the missing 4-momentum by requiring vanishing missing mass squared~\cite{CLEOmxq}. Although high reconstruction efficiency is obtained only bad $M_X$ resolution is obtainable since the B momentum in CMS of the $\Upsilon(4S)$ must be neglected  and as a result signal/noise ratio is poor.
Second method is a full-reconstruction method used by BaBar where one of two B mesons is fully reconstructed. In this case good $M_X$ resolution is obtained, but the reconstruction efficiency is poor like a few times $10^{-3}$~\cite{BaBar}. The third method is a  novel $X_u$-reconstruction method of Belle~\cite{Bellemxq} based on a combination of neutrino reconstruction and a technique called simulated annealing~\cite{anneal} to separate the two $B$ meson decays.
This method allows us to measure $M_X$ and $q^2$ with good efficiency so that it  achieves good statistical precision and small theoretical uncertainty with a modest integrated luminosity.
We report here the first result with simultaneous requirements on $M_X$ and $q^2$~\cite{Bellemxq}.
This analysis is based on $78.1\ {\rm fb}^{-1}$  data, corresponding to 85 million $B \bar{B}$ pairs, taken at the $\Upsilon(4S)$
resonance, and $8.8\ {\rm fb}^{-1}$ taken at an energy 60~MeV below the resonance, by the Belle detector
\cite{Belle} at the energy-asymmetric  $e^+ e^-$ collider KEKB~\cite{KEKB}.

\section{Experimental Procedure}
\subsection{Neutrino Reconstruction}\label{subsec:neutrino}
\begin{figure}[htb]
\includegraphics[width=8.0cm,height=4.0cm]{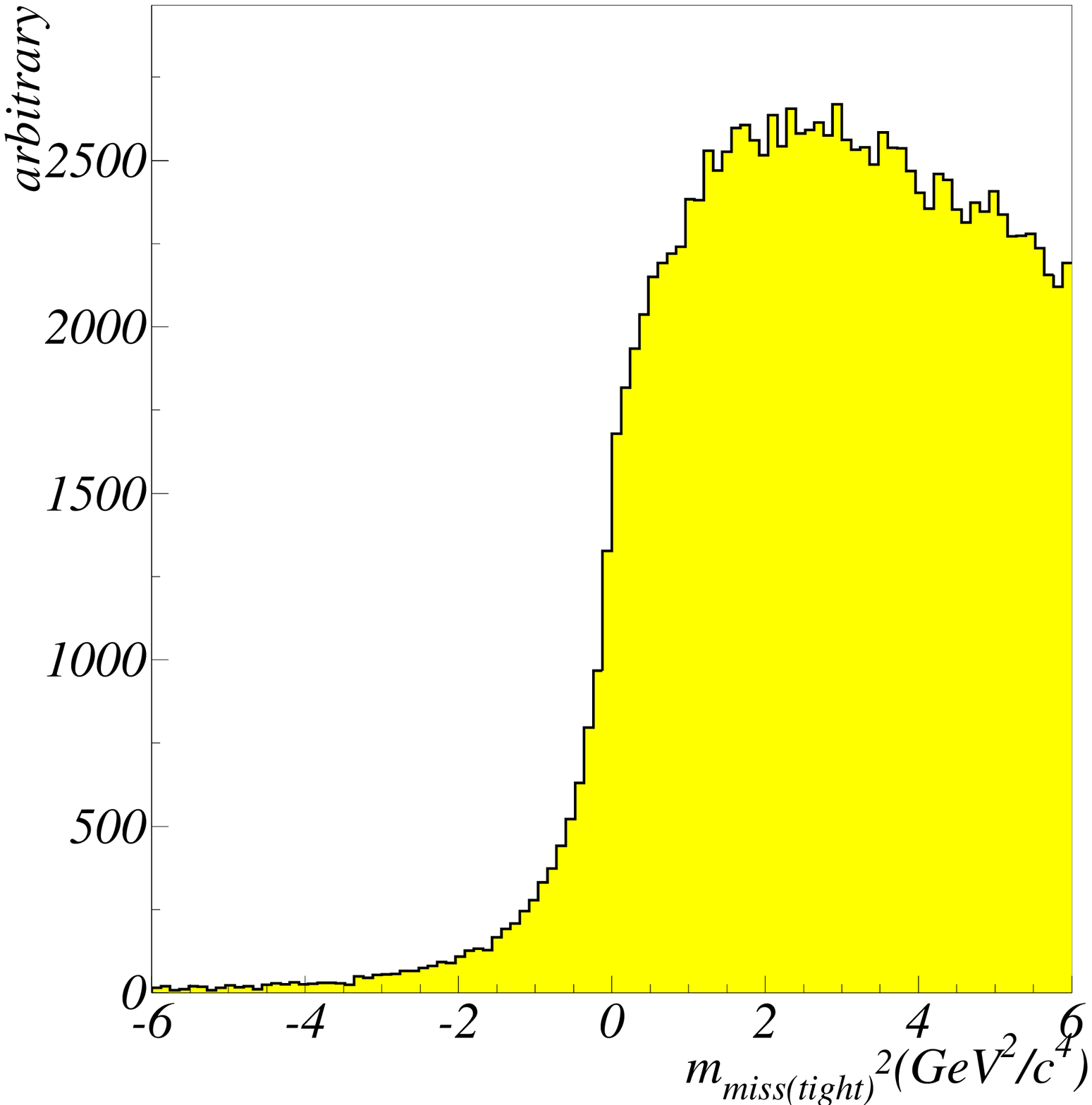}
\includegraphics[width=8.0cm,height=4.0cm]{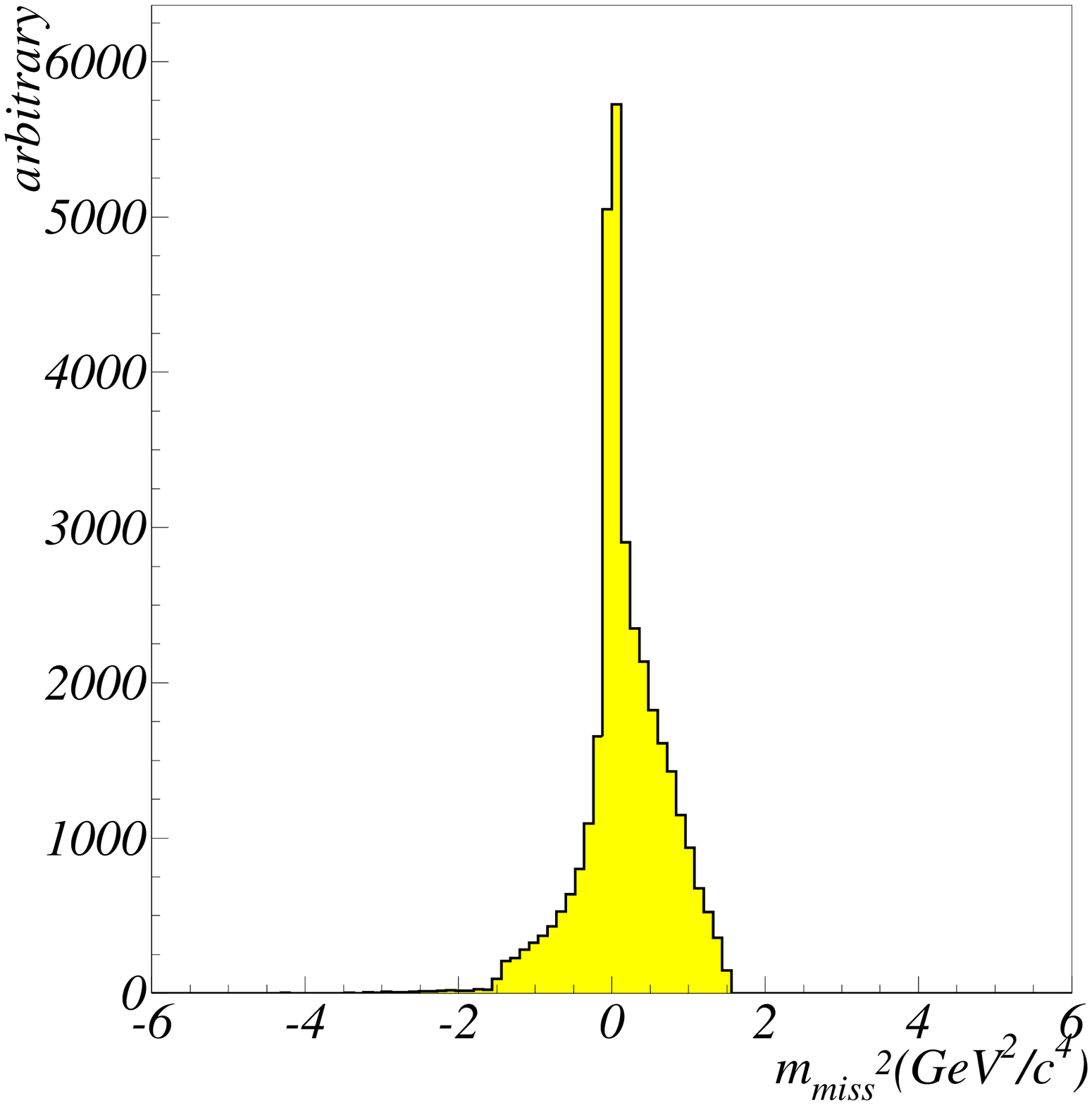}
\caption{$\mathit{MM}^2$ distributions. The left and right plots correspond
to the case with good quality tracks and the case including poor quality tracks, respectively.}
\label{mm2}
\end{figure}

We select hadronic events containing one lepton candidate having momentum above 1.2~GeV/$c$ in CMS of the $\Upsilon(4S)$. 
We exclude events containing
additional lepton candidates. 
The neutrino is reconstructed
from the missing 4-momentum in the event
($\vec{p}_{\nu} \equiv \vec{p}_{\Upsilon(4S)} - \Sigma_i \vec{p}_i$,
$E_{\nu} \equiv E_{\Upsilon(4S)} - \Sigma_i E_i$). 
The net observed momentum ($\Sigma_i \vec{p}_i$) and energy ($\Sigma_i E_i$) are calculated using particles surviving track quality cuts.
We then compute the missing mass of the event, defined as
$\mathit{MM}^2\equiv E_{\nu}^2/c^4-|\vec{p}_{\nu}|^2/c^2$, where the sign of $E_{\nu}^2$ is reversed when $E_{\nu} < 0$.
The result is shown in the left plot of Figure~\ref{mm2}. 
We require  $-1.5\ {\rm GeV}^2/c^4<\mathit{MM}^2<1.5\ {\rm GeV}^2/c^4$
to suppress events with missing particles 
and with particles removed due to poor reconstruction quality.
For events that pass this requirement, 
we add back tracks and clusters rejected earlier due to reconstruction quality, selecting the combination that
gives the smallest value of $|\mathit{MM}^2|$.
The result is shown in the right plot of Figure~\ref{mm2}. 
This determines the set of particles that are used in the subsequent analysis.
Events are further required to have a net
charge of 0 or $\pm 1$, and a polar angle for the missing momentum
within the barrel region ($32^{\circ} < \theta < 128^{\circ}$).
To suppress beam-gas events, we demand that the  net charge of all proton candidates be zero. 
Requiring that the
cosine of the angle of  $K^0_L$ candidates with respect to the
 missing momentum be less than 0.8 rejects events where the neutrino candidate
  is actually a $K^0_L$ meson. 

\subsection{Simulated Annealing}\label{subsec:annealing}
We then seek the most likely combination of particles belonging to $X \ell \nu$, the remainder
being from the associated $B$-meson ($B_{\rm opp}$).
Six discriminant variables are used: the momentum, energy, and polar angle of $B_{\rm opp}$ ($p_B^*$, $E_B^*$, $\cos \theta_B^*$) in the CMS, 
its charge
multiplicity ($N_{\mathrm{ch}}$) and net charge times the lepton charge ($Q_{B}\times Q_{\ell} $), and the missing-mass squared
recalculated with the energy and mass of $B_{\rm opp}$ constrained to the known values
($\mathit{MM}^2_{X \ell \nu}$).
Using Monte Carlo (MC) simulation events for $\Upsilon(4S) \to B \bar{B}$
where at least one $B$ decays into $X \ell \nu$, 
we determine  probability density functions (PDFs) 
for correct $X \ell \nu$ combinations
and for random $X \ell \nu$ combinations. 
From the PDFs we calculate two likelihoods,
${\cal L}({\rm correct})$ and ${\cal L}({\rm random})$.
The most likely candidate combination in each event  is found by minimizing the parameter 
$ W \equiv {\cal L}({\rm random})/({\cal L}({\rm random}) + 
    {\cal L}({\rm correct}))$.

\begin{figure}[htb]
\includegraphics[width=14cm,height=7.0cm]{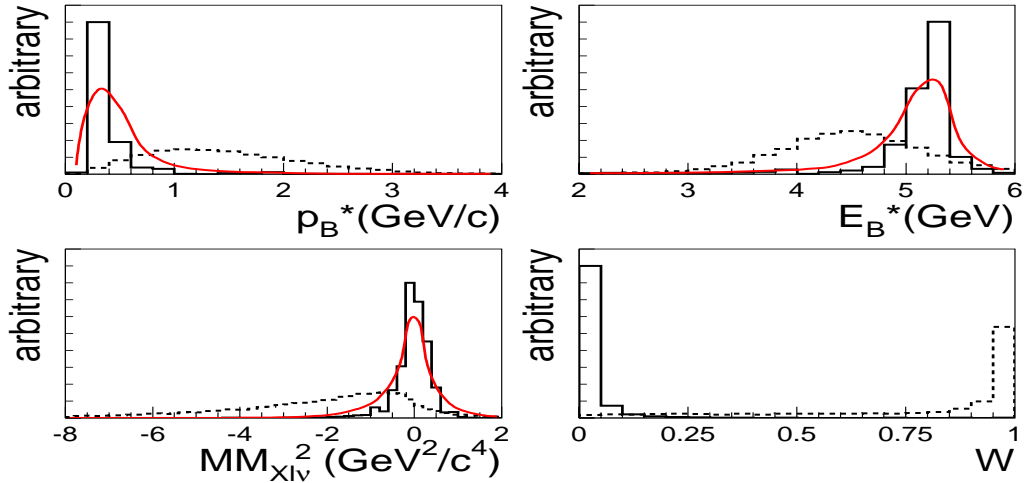}
\caption{Distributions for 3 discriminants and $W$ before (dashed histogram)  
and after (solid histogram) simulated annealing,  for real data. MC distributions for 3 discriminants for correct combination of particles (solid curve). }
\label{discriminant}
\end{figure}

To minimize $W$, we have developed an approximate iterative algorithm based on simulated annealing.
We start from the initial candidate for $X \ell \nu$ that consists of the lepton and neutrino plus approximately one third of the remaining particles, selected randomly.
We move a random particle (other than the lepton or neutrino) between the $X$ and $B_{\rm opp}$ sides in an iterative way, where in one iteration we cross all particles at least once, and search for the combination 
that gives the minimum 
$W$ with 50 iterations. During the iteration process we take special care to reduce the chance of convergence to a local minimum of $W$.
For instance, after every fifth iteration we compare
all combinations that can be constructed by crossing one particle and 
use the combination that gives the {\it largest} value of $W$ to seed a new cycle.
We repeat this iteration process  10 times, starting each time with a different initial candidate, and select the case  with the smallest $W$.
Figure~\ref{discriminant} shows the
distributions in three of the six discriminant variables and $W$,  before and after simulated annealing. 
Also shown are the 
distributions for the correct combination in signal MC events.   
The final candidate is required to satisfy: 
i) $W < 0.1$, ii) $5.1 < E_B^* < 5.4$~GeV, iii) $0.25 < p_B^* <
0.42$~GeV/$c$, iv) $-2 < Q_{\ell} \times Q_{B} < +1$, and v) $-0.2 < \mathit{MM}^2_{X \ell \nu} < 0.4$~GeV$^2$/$c^4$. 
Contamination from the continuum is reduced  by demanding 
$|\cos \theta_{B\ell}|<0.8$, where $\theta_{B\ell}$ is the
angle between the thrust axis of $B_{\rm opp}$ and the lepton momentum. 

\subsection{Validity Check}\label{subsec:validity}
The validity of the  method is checked with two data samples, one containing 38,600 fully-reconstructed $B \to D^* \ell \nu$ decays and the other containing
 84,100 $B \to J/\psi X, J/\psi\to\ell^+\ell^-$ decays.
For the $J/\psi X$ sample we treat one of the two leptons from $J/\psi$
as a neutrino, to emulate $B\to X \ell \nu$.
%Corresponding MC events are generated with the QQ event generator~\cite{QQ}
% and the detector response is simulated using Geant~3~\cite{geant3}. 
The $M_X$ distribution for the $D^* \ell \nu$ sample is peaked at the $D^*$ mass and  the $q^2$ distribution for the $J/\psi X$ sample is peaked at $J/\psi$ mass squared, as expected.
The shapes are in good agreement between data and MC. 
These results verify that the simulated annealing method works as expected. 
We use the efficiency 
ratio $r_{\rm{eff}}=0.891\pm0.043$ between the data and MC for calibration
of signal efficiency by MC. 

\subsection{Background Subtraction}\label{subsec:background}
\begin{figure} [htb]
\begin{minipage}{7.5cm}
\includegraphics[width=8.0cm,height=7.0cm]{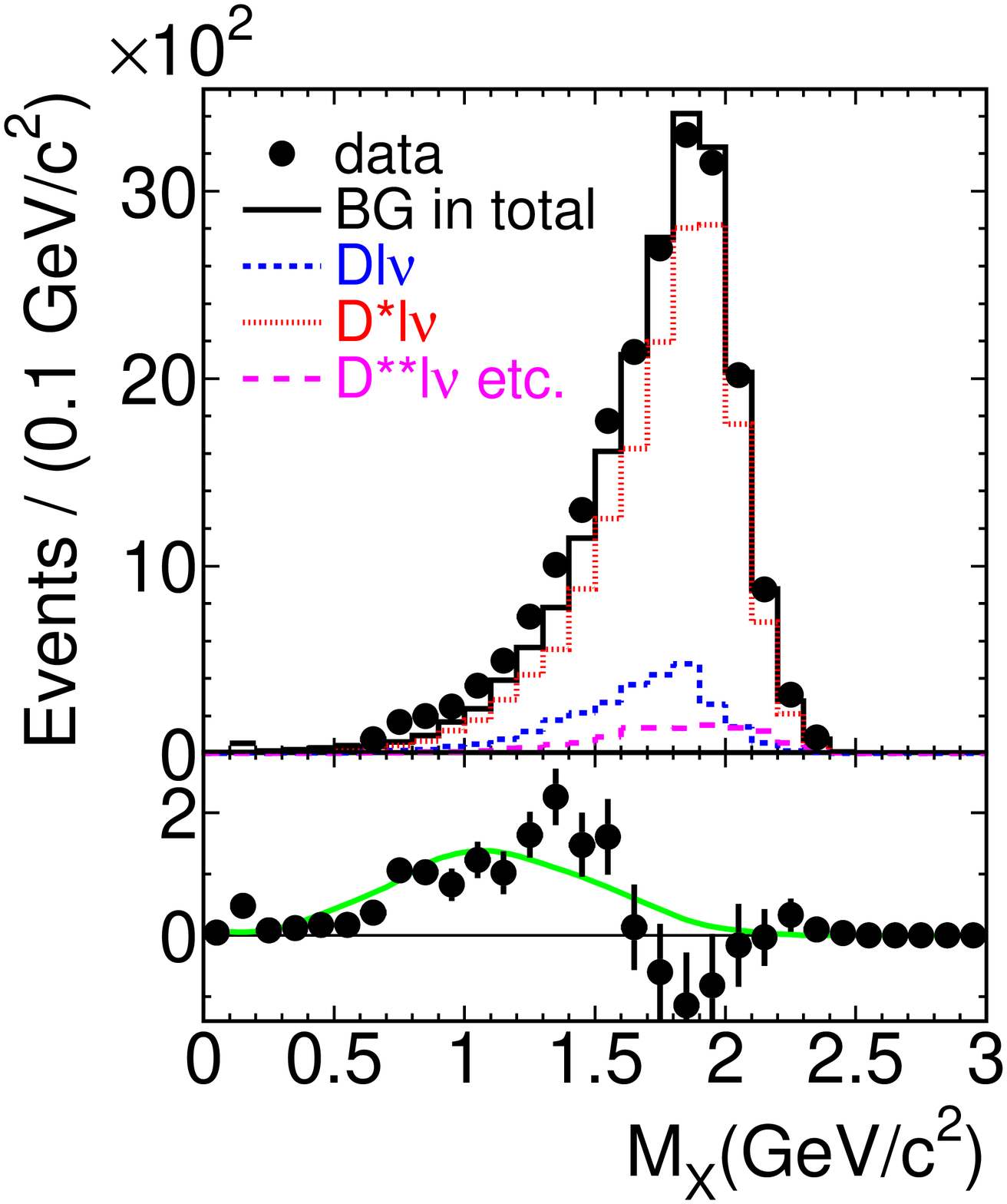}
\end{minipage}
\begin{minipage}{7.5cm}
\includegraphics[width=8.0cm,height=7.0cm]{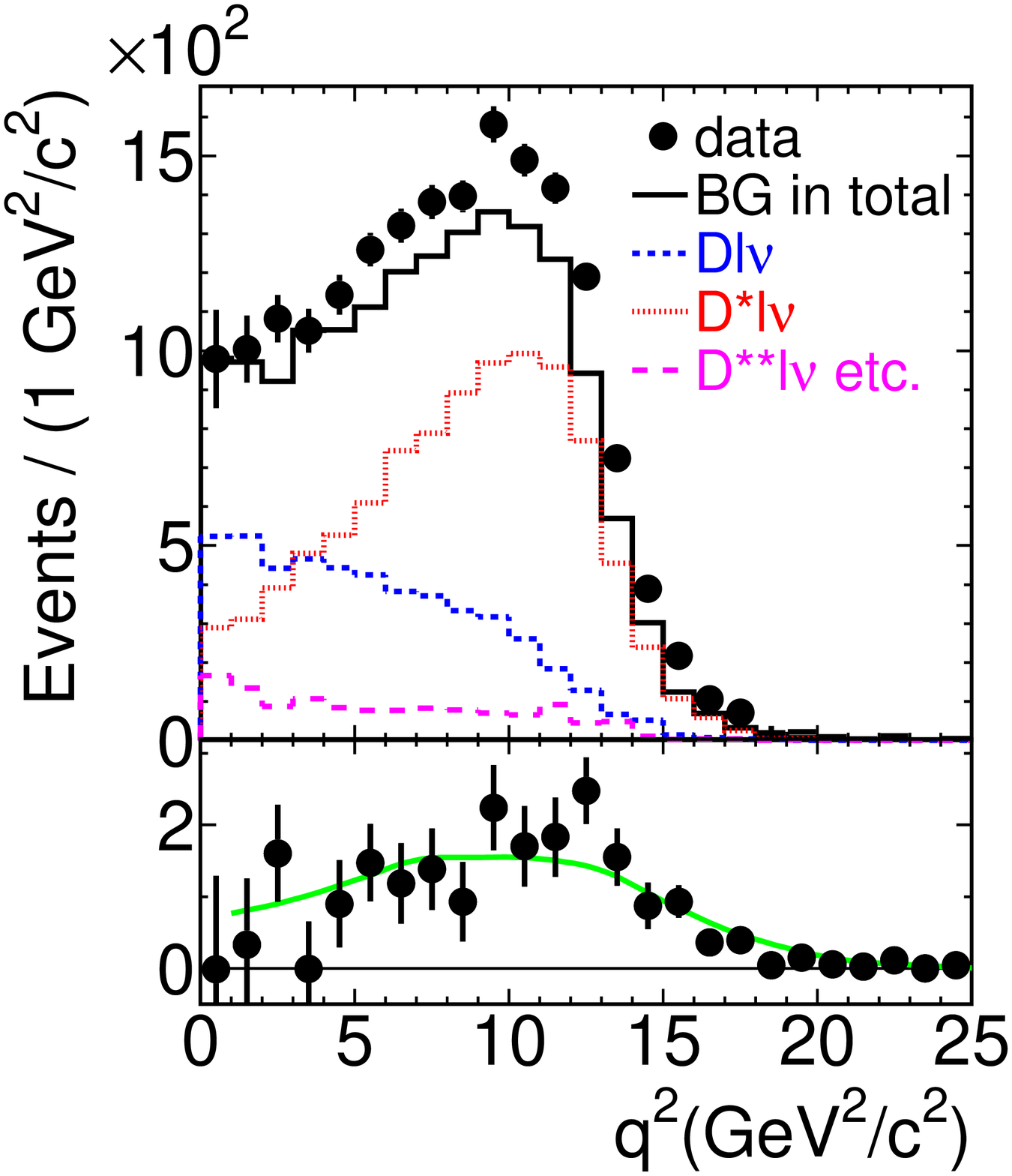}
\end{minipage}
\caption{(a) $M_X$ distribution for $q^2 > 8.0~\mathrm{GeV^2}/c^2$. 
(b) $q^2$ distribution for  $M_X<1.7\ \mathrm{GeV}/c^2$.
Points are the data and histograms are backgrounds from 
$D^* \ell \nu$ (dotted), $D \ell \nu$ (short dashed), others (long dashed), and total background contribution (solid). Lower plots show the data after background subtraction.
Solid curves show the inclusive MC predictions for $ B \to X_u \ell \nu$.}
\label{q2mx}
\end{figure}

We observe 8910 events in the $X_u \ell \nu$ ``signal'' region, defined as $M_X < 1.7 {\rm{GeV}}/c^2$ and $q^2 > 8.0 {\rm{GeV}}^2/c^2$.
These consist of semileptonic decays, $B\to X_{c,u} \ell
\nu$, other $B\bar B$ background, and residual continuum events.
The continuum contribution is estimated from off-resonance data to be $251\pm48$ events and is subtracted directly from the analyzed distributions.
The contributions from $B\to X_c \ell \nu$ and other $B\bar B$ backgrounds are estimated via MC in the ``background'' region $ M_X > 1.8\ {\mathrm{GeV}}/c^2$,  where  
$X_c \ell \nu$ dominates, and extrapolated to the signal region.
 We estimate them by fitting the $M_X$ and $q^2$ distributions from MC events  to those from the data using a two-dimensional $\chi^2$ fit method.
Contributions to  $X_c \ell \nu$ come from $D^{(*)} \ell \nu$, $D^{**} \ell \nu$,  and $D^{(*)} \pi \ell \nu$. 
Their branching fractions are floated in the fit.
The total rate for other $B\bar B$ backgrounds is also floated.
The small $X_u \ell \nu$ contribution is estimated iteratively.
The obtained branching fractions are consistent  with the PDG values~\cite{PDG2002}. 
The $B \bar{B}$ background in the signal region is estimated to be $7283\pm130\pm63$ events, where the first and second errors come from  fit and MC statistics, respectively.
The upper plots in Figure~\ref{q2mx}  show the $M_X$ distribution for $q^2 > 8.0\  {\rm{GeV^2}}/c^2$ and the $q^2$ distribution for  $M_X<1.7\ {\mathrm{GeV}}/c^2$ after continuum subtraction.
After subtracting the $B\bar B$ backgrounds, we obtain
 distributions for the $X_u \ell \nu$ signal,
shown in the lower plots. 
The net signal is $N_{\rm{obs}}$ = 1376$ \pm 167$ events where the error is statistical only.

\section{Extraction of Partial Branching Fraction}

In order to extract the partial branching fraction $\Delta \mathcal{B}$ for $B \to X_u \ell \nu$ in the signal region,
a Monte Carlo simulation is used to convert  $N_{\rm{obs}}$ to the  true number of signal events produced in this region, $N_{\rm{true}}$, and to estimate the efficiency for these events to be observed anywhere, $\epsilon_{\rm{signal}}$. In the MC simulation, $B \to X_u \ell \nu$ decays are simulated based on the prescription of Fazio and Neubert~\cite{fn}.
For two parameters therein we use $m_b = 4.80\pm0.12\, {\rm{GeV}}/c^2$ and $\mu_{\pi}^2 =0.30\pm0.11\,{\rm{GeV}}^2/c^2$ derived from the CLEO measurements of the hadronic mass moments in inclusive $B \to X_c \ell \nu$ and photon energy spectrum in $B \to X_s\gamma$~\cite{lambda}.
% The MC events are generated with the EvtGen generator~\cite{evtgen}.
$ N_{\rm{true}}$ is estimated by
$ N_{\rm{true}} = N_{\rm{obs}} \times F $.
We find $F=0.938$, and thus
$N_{\rm{true}} = 1291\pm 157$. 
The efficiency   $\epsilon_{\rm{signal}}$ is predicted to be 0.578\%.
We  determine $\Delta \mathcal{B}$  by
$0.5\times N_{\rm{true}}/(\epsilon_{\rm{signal}}\times r_{\rm{eff}})/(2N_B)$,
where $r_{\rm{eff}}$ is the efficiency correction factor described earlier, $N_B$ is the number of $B\bar{B}$ events
and  the factor  0.5 is needed to take into account the electron and muon
data:
\begin{eqnarray}  
 \Delta\mathcal{B} = (7.37\pm0.89\pm1.12 \pm 0.55 \pm0.24)\times10^{-4}. \nonumber
\end{eqnarray}
The errors are statistical, systematic, from 
$B \to X_c \ell \nu$ model dependence, and $B \to X_u \ell \nu$ model dependence, respectively.
Main source of  systematic uncertainty is the uncertainty in the $B \bar{B}$ background estimation arisen from distortion of the $M_X$ and $q^2$ distributions due to imperfect detector simulation.
The  model dependence of $X_c \ell \nu$ is estimated to be 7.4{\%} by 
 varying  
the $D_1 \ell \nu$ plus $D^*_2 \ell \nu$ fraction in the $D^{**}\ell \nu$
by 25{\%} and by varying  the slope parameters of the form factors for $D \ell \nu$ and $D^* \ell \nu$,  $\rho_D^2=1.19\pm0.19$ and $\rho^2=1.51\pm0.13$ ~\cite{PDG2002}, within their errors.
$B \to X_u \ell \nu$  model dependence  (3.4{\%}) is estimated by varying the parameters of the inclusive model within their errors and by comparing to a
simulation with a full exclusive implementation of the ISGW2 model~\cite{ISGW2}.

\section{Extraction of $|V_{ub}|$}

As mentioned before $\Delta\mathcal{B}(B \to X_u \ell \nu)$ is related to $|V_{ub}|$~\cite{bau,HLM,ligeti},
\begin{eqnarray}  
|V_{ub}| = 0.00444 \left(  \frac  {\Delta \mathcal{B}(B \rightarrow X_u \ell \nu )}
                       {0.002\times 1.21G(q^2_{\rm{cut}},m_{\rm{cut}})} 
                             \frac{1.55\mathrm{ps}}
                                  {\tau_B}   
                              \right)^{1/2}
\end{eqnarray}
where 
$1.21G(q^2_{\rm{cut}},m_{\rm{cut}})= f_u \times \left( \frac {m_b^{\rm{1S}}}{4.7{\rm{GeV}}/c^2}\right)^5 $,  $f_u$ represents the fraction of events with $q^2>q^2_{\rm{cut}}$ and $M_X<m_{\rm{cut}}$.
%,and
%$m_b^{\rm{1S}}$ is one-half of the perturbative contribution to the mass of the $\Upsilon(1S)$.
$G(q^2_{\rm{cut}},m_{\rm{cut}})$ is calculated 
to ${\cal O}(\alpha_s^2)$ and ${\cal O}(1/m_b^2)$ in Ref.4, including the effect of the Fermi motion of the $b$ quark,
 which is expressed in terms of $m_b^{\rm{1S}}$.
We use $m_b^{\rm{1S}}=4.70\pm0.12\,{\rm{GeV}}/c^2$~\cite{bau,mb1s}, which gives
$G(q^2_{\rm cut},m_{\rm cut})=0.268$ \cite{bau,ligeti}.
 The theoretical uncertainty on $|V_{ub}|$ is determined only by the uncertainty on $G(q^2_{\rm{cut}},m_{\rm{cut}})$.
The uncertainty on  $G(q^2_{\rm{cut}},m_{\rm{cut}})$,  in total 25$\%$,  consists of 6{\%} for perturbative, 8{\%} for nonperturbative terms (dominated by the weak annihilation contribution),  and 23{\%} from  the uncertainty on $m_b^{\rm{1S}}$~\cite{bau,duality}.
The 23\% error contains 10\% for 
$f_u$ and 13\% for  $(m_b^{\rm{1S}})^5$. 
These uncertainties are positively correlated, so we add them
linearly, whereas they have been given separately in conventional
analyses.
Using $\tau_B =1.604\pm0.012$ ps~\cite{PDG2002},
we obtain 
\begin{eqnarray}  
|V_{ub}| &= & (4.66 \pm 0.28 \pm0.35 \pm 0.17 \pm 0.08 \pm  0.58) \times
10^{-3} \nonumber
\end{eqnarray}  
where the errors are statistical, systematic,  $b \to c$ model dependence, 
$b \to u$ model dependence, and theoretical uncertainty for OPE, respectively.

\section{Summary}
We have performed the first measurement of $|V_{ub}|$ with simultaneous requirements on $M_X$ and $q^2$ using a novel $X_u$-reconstruction method. The result of $|V_{ub}| = (4.66 \pm 0.76) \times 10^{-3}$ is consistent with the previous inclusive measurements~\cite{CLEOendpt,BaBar,LEP}  and the  total error  is comparable with those of the  previous measurements on $\Upsilon(4S)$~\cite{CLEOendpt,BaBar}.
Due to simultaneous requirements on $M_X$ and $q^2$, the extrapolation error is much smaller than those  of the previous measurements on $\Upsilon(4S)$~\cite{CLEOendpt,BaBar}.

%\begin{figure}
%\rule{5cm}{0.2mm}\hfill\rule{5cm}{0.2mm}
%\vskip 2.5cm
%\rule{5cm}{0.2mm}\hfill\rule{5cm}{0.2mm}
%%\psfig{figure=filename.ps,height=1.5in}
%\caption{Radiative (off-shell, off-page and out-to-lunch) SUSY Higglets.
%\label{fig:radish}}
%\end{figure}

\section*{Acknowledgments}
I would like to thank the organizers of the conference for arranging a stimulating conference in a pleasant atmosphere. I am grateful to H.~Kakuno for assistance in preparing this report.


\begin{thebibliography}{99}
\bibitem{duality} The uncertainty due to the possible quark-hadron duality violation is not included in this estimation.
\end{thebibliography}
\end{document}